# Higher-order photonic topological states in surface-wave photonic crystals

Li Zhang[1,2,#], Yihao Yang[3,4,#, *], Pengfei Qin[1,2], Qiaolu Chen[1,2], Fei Gao[1,2], Erping Li[1,2], Jian-Hua Jiang[5,*], Baile Zhang[3,4,], and Hongsheng Chen[1,2,*]

[1]State Key Laboratory of Modern Optical Instrumentation, College of Information Science and Electronic Engineering, Zhejiang University, Hangzhou 310027, China.

[2]Key Laboratory of Advanced Micro/Nano Electronic Devices & Smart Systems of Zhejiang, The Electromagnetics Academy at Zhejiang University, Zhejiang University, Hangzhou 310027, China.

[3]Division of Physics and Applied Physics, School of Physical and Mathematical Sciences, Nanyang Technological University, 21 Nanyang Link, Singapore 637371, Singapore.

[4]Centre for Disruptive Photonic Technologies, The Photonics Institute, Nanyang Technological University, 50 Nanyang Avenue, Singapore 639798, Singapore.

[5]School of Physical Science and Technology, and Collaborative Innovation Center of Suzhou Nano Science and Technology, Soochow University, 1 Shizi Street, Suzhou 215006, China.

# These authors contributed equally to this work.

[*] yang.yihao@ntu.edu.sg (Yihao Yang); jianhuajiang@suda.edu.cn (Jian-Hua Jiang); hansomchen@zju.edu.cn (Hongsheng Chen).

## ABSTRACT

Photonic topological states have revolutionized our understanding on the propagation and scattering of light. Recent discovery of higher-order photonic topological insulators opens an emergent horizon for zero-dimensional topological corner states. However, the previous realizations of higher-order photonic topological insulators suffer from either a limited operational frequency range due to the lumped components involved or a bulky structure with a large footprint, which are unfavorable for future integrated photonics. To overcome these limitations, we hereby experimentally demonstrate a planar surface-wave photonic crystal realization of two-dimensional higher-order topological insulators. The surface-wave photonic crystals exhibit a very large bulk bandgap (a bandwidth of 28%) due to multiple Bragg scatterings and host one-dimensional gapped edge states described by massive Dirac equations. The topology of those higher-dimensional photonic bands leads to the emergence of zero-dimensional corner states, which provide a route toward robust cavity modes for scalable, integrated photonic chips and an interface for the control of light-matter interaction.

## INTRODUCTION

Photonic topological insulators (PTIs) [1-19] host unprecedented edge states such as the chiral or helical edge states in two-dimensional (2D) PTIs [4,9,10,12,13,18] and the Dirac-fermion-like surface states in three-dimensional PTIs [19,20]. The topologically protected edge states can lead to important applications such as high-transmittance waveguides [2,10,14,16], robust photonic delay lines [3,13], topological wave partition [18], robust photonic transport on non-planar surfaces [19], topological lasers [21,22], and topological quantum interfaces [23].

The quest for topological zero-dimensional (0D) cavity modes in 2D photonic systems [24-26], which may provide a route toward the build-up of robust and scalable integrated photonic circuits, was unsuccessful until very recently [27,28]. Such an achievement is realized using the higher-order topological insulators [27-37]. Unlike the conventional $D$-dimensional topological insulators which lead to ($D$-1)-dimensional topological edge states, a $D$-dimensional higher-order topological insulator gives rise to ($D$-2)-dimensional (or even lower-dimensional) topological boundary states, in addition to the ($D$-1)-dimensional topological edge states, beyond the conventional bulk-boundary correspondence. Through the concept of higher-order topological insulators, it has been demonstrated that topological 0D corner states can emerge in microwave circuits with lumped components [28] and coupled optical waveguides [27]. However, the microwave circuits with lumped components can work at low frequencies (normally up to several GHz), and the coupled optical waveguides have lattice constants much larger than the wavelength ($a \approx 40\lambda$) and a small bandgap (2%). These photonic systems are hence unfavorable for future integrated photonics.

Here, we demonstrate a second-order PTI which realizes photonic quantum spin Hall effect through topological crystalline order and gives rise to gapped edge states and topological corner states in surface-wave photonic crystals (PCs) with $a \approx 0.5\lambda$. The surface-wave PC design provides a very large photonic bandgap (a bandwidth of 28% at the M point below the light-line) as induced by multiple Bragg scatterings. The symmetry-guided approach enables us to tune the photonic bandgap and band topology in both the bulk and the edges, simultaneously. In particular, the orthogonal edges in a finite square structure are described by 1D massive Dirac equations with tunable Dirac masses, which leads to topological localization of light at the corners connecting the orthogonal edges through the Jackiw-Rebbi mechanism [38,39] (see Fig. 1a). The gapped topological edge states, as well as the topological corner states within the edge bandgap, are directly observed and characterized in our experiments. Compared with the coupled optical waveguide arrays which have a large lattice constant ($a \approx 40\lambda$) and small photonic bandgap (2%) and the microwave circuits with lumped components working at low frequencies, the present surface-wave PCs with a small lattice constant can lead to pronounced advantages [24-26] such as large photonic bandgap, miniature structures, and the compatibility for the integration of optoelectronic devices. The 0D topological corner states in PCs provide an efficient way for scalable integration of cavity modes with identical and robust frequency for integrated photonics and optoelectronics.

## RESULTS

As depicted in Fig. 1a, the design of the surface-wave PC consists of metallic patterns on both sides of a dielectric substrate in a square lattice with a lattice constant $a$ =12 mm. Each unit cell is composed of four metallic rectangle patterns on one side, with a width $w$ =1.92 mm and a length $l$ =5.04 mm. The rotation angle $\theta$ of the metallic rectangles (see the upper-right inset of Fig. 1a) can be tuned, which controls the photonic bands and band topology. Note that here we choose a unit-cell which doubles the primitive unit-cell. However, the chosen unit-cell is the smallest unit-cell that is compatible with the supercell structure for the edge and corner states [38] studied in this work. In experiments, the designed surface-wave PCs are fabricated by printing double-sided 0.035-mm-thick copper cladding onto 2-mm-thick F4B printed circuit boards (relative permittivity 3). The 3D electromagnetic field profiles are presented in details in the Supplementary Information.

The surface-wave PC has a large topological bandgap. For $\theta$=45°, the frequency gap below the light-line ranges from 11.5 GHz to 15.2 GHz, leading to a bandwidth of 28% with a mid-gap frequency of 13.3 GHz. The latter corresponds to a lattice-constant/wavelength ratio of $a/\lambda \sim 0.5$, indicating the subwavelength nature of the surface-wave PC.

The structure of the surface-wave PC has glide-reflection symmetries in both the $x$ direction $g_x$ ={$m_x|\tau_y$} and the $y$ direction $g_y$ ={$m_y|\tau_x$}, where $m_x\psi(x,y,z)$ =$\psi(-x,y,z)$, $\tau_y\psi(x,y,z)$ =$\psi(x,y+a/2,z)$, $m_y\psi(x,y,z)$ =$\psi(x,-y,z)$ and $\tau_x\psi(x,y,z)$ =$\psi(x+a/2,y,z)$. Combining the glide-reflection and time-reversal symmetries, Kramer-like double degeneracies exist at the boundaries of the Brillouin zone when $k_x$ =$\pi/a$ or $k_y$ =$\pi/a$ (the MX and MY lines) [38,40,41], as shown in Fig. 1b. When $\theta$ =0° or 90°, a pair of doubly-degenerate bands cross each other and form a four-fold Dirac degeneracy at the M point in the Brillouin zone at a frequency of 12.97 GHz (see the black curves in Fig. 1b), which can also be explained by the Brillouin zone folding mechanism (See Supplementary Information). When $\theta$ deviates from 0° or 90°, the four-fold Dirac degeneracy splits into two pairs of doubly-degenerate bands, and a photonic bandgap appears. For instance, the photonic band structure for $\theta$ =45° is shown in Fig. 1b (the red and green curves for the "conduction" and "valence" bands, respectively).

The evolution of the band edges at the M point with the rotation angle $\theta$ is shown in Fig. 1c. These band edges consist of two pairs of photonic states of opposite parities (see Fig. 1d): the even-parity, quadrupole-like modes ($d_{x^2-y^2}$ and $d_{xy}$) and the odd-parity, dipole-like modes ($p_x$ and $p_y$). When $\theta$ goes through 0° or 90°, the photonic bandgap experiences a parity switch, indicating a topological phase transition [38]. There are two topologically distinct phases: the normal insulator phase (NIP) and the topological insulator phase (TIP). The latter is a topological crystalline insulator which mimics the quantum spin Hall effect in photonic systems, as shown in the following.

We now investigate the topological edge states emerging at the interface between the NIP and TIP PCs. Since the PCs may not have $C_4$ rotation symmetry, the edge states for the interfaces along the *x* and *y* directions (shortened as the *x*- and *y*-interfaces) are generally different. We study the photonic edge states at both the *x*- and *y*-interfaces for two situations: First, for the PCs with $\theta_1$ =-25° (NIP) and $\theta_2$ =25° (TIP) (see Fig. 2a); Second, for the PCs with $\theta_1$ =-25° (NIP) and $\theta_2$ =50° (TIP) (see Fig. 2b). Fig. 2a demonstrates a case where the *x*- and *y*-interfaces preserve the glide symmetries (this holds whenever $\theta_2$ =-$\theta_1$), while Fig. 2b demonstrates a case where such symmetries are broken on those edges.

Fig. 2a illustrates that the photonic edge states are time-reversal symmetric helical edge states exhibiting pseudospin-wavevector locking, where the pseudospins are emulated by the photonic orbital angular momentum as indicated by the winding of the Poynting vectors in the electromagnetic field profiles (see Supplementary Information for more details). These pseudo-spins confirm that we have realized a photonic analog of the quantum spin Hall effect. However, due to their bosonic nature, the photonic edge states are not protected by the time-reversal symmetry. Consequently, the edge states can be gapped whenever the glide symmetries are absent at the edges (as exampled in Fig. 2b), while the photonic edge states are described by 1D massive Dirac equations. These gapped edge states can be described by the Hamiltonians $H_\alpha = v_\alpha (k_\alpha - \frac{\pi}{a})\sigma_z + m_\alpha \sigma_y$ with $\alpha = x, y$ for the *x*- and *y*-interfaces, where $v_\alpha$ is the group velocity of the edge states;

$\sigma_z = 1$ and -1 represent the pseudospin-up and -down modes. The Dirac masses $m_\alpha$ are determined by half of the frequency difference between the even and odd modes at $k_\alpha = \frac{\pi}{a}$ ($\alpha = x, y$). The topology of the 1D photonic edge states is characterized by the sign of the Dirac masses. Parity switch in the edge states thus signals the topological transition in the edge states.

The Dirac masses of the edge states at *x*- and *y*-interfaces, $m_x$ and $m_y$, are generally different and can be controlled by the rotation angles $\theta_1$ and $\theta_2$. In Figs. 2c and 2d, we show the Dirac mass as a function of $\theta_2$ for 0°≤ $\theta_2$ ≤ 90°, with $\theta_1$ fixed at -25°. There are three topological transition points for the edge states, $\theta_2$ =0°, 25°, and 90°, in those phase diagrams. The transition points $\theta_2$ =0° and 90° are associated with the topological transitions of the bulk photonic bands. In contrast, the transition point $\theta_2$ =25°, with both $m_x$ and $m_y$ equal to zero, is solely due to the edge (as shown in Fig. 2c), resulting from the restoration of the glide symmetries at the two interfaces. Interestingly, despite the topological transition at $\theta_2$ =25°, the signs of $m_x$ and $m_y$ remain opposite before and after the transition, which manifests the stability of the higher-order band topology.

In the supercell structure illustrated in Fig. 1a, the opposite signs of $m_x$ and $m_y$ lead to the formation of 0D photonic states localized at the four corners connecting the *x*- and *y*-interfaces, due to the Jackiw-Rebbi mechanism [39]. Hence, in our system the 2D bulk topology results in the topological 1D edge states, while the topology of the gapped 1D edge states leads to the 0D corner states. This manifestation of bulk-edge correspondence in a hierarchy of dimensions reveals a hallmark feature of the second-order topology.

In the experiments, we fabricate two different square-shaped super-structures with both NIP and TIP PCs. The first sample consists of the PC with $\theta_2$ =25° (TIP) surrounded by the PC with $\theta_1$ =-25° (NIP) (see Fig. 3a). The edge states are excited by a dipole source near the bottom of the sample located at the position labeled by the red star in Fig. 3a. The response at the *x*-edge (*y*-edge) is detected by a probe located at the blue (green) dot. The response of the bulk is measured by another probe at the center of the TIP PC (see details in the Supplementary Information). The measured (normalized) $|H_z|^2$ field intensities at those detection positions are shown in Fig. 3b. One can see that the transmission of the bulk (the grey region) are very low in the frequency range from 12.5 GHz to 14 GHz, in consistency with the bulk bandgap. Within this frequency range, the responses for the edges along both the *x*- (the blue region) and *y*- (the green region) interfaces are much stronger than the bulk, indicating the edge states within the bulk bandgap. Both the *x*- and *y*-edges have continuous (gapless) responses, indicating their gapless spectrum, in agreement with the edge dispersion in Fig. 2a. The electromagnetic field profile measured directly by a near-field scanning system (see Supplementary Information) is shown in Fig. 3c. The measured field profile and the corresponding simulation (Fig. 3d) indicate light flow along the edges since the exciting frequency 13.24 GHz is in the bulk bandgap.

We further measure a sample consisting of a TIP PC with $\theta_2 = 50°$ surrounded by a NIP PC with $\theta_1 = -25°$, as illustrated in Fig. 4a. We measure the responses of the edge and bulk states in similar means as in Fig. 3. The results in Fig. 4b reveal that both the *x*- and *y*-edges have gapped photonic spectrum within the bulk bandgap (i.e., from 12.5 GHz to 14 GHz). We also measure the response of the corner by placing a probe with two unit-cell distance away from the dipole source. The corner response clearly indicates a strong and sharp peak at 12.71 GHz within the spectral gaps of both the *x*- and *y*-edges (their common frequency gap ranges from 12.6 GHz to 12.8 GHz). Such a sharp resonance indicates the emergence of the topological corner modes. The spectral fingerprints with edge states in the bulk bandgap and corner states in the edge bandgap is a smoking-gun feature of second-order topological insulators.

The corner resonance is further studied by measuring its electromagnetic profile using the near-field scanning method. The measurement results, shown in Fig. 4c, is consistent with the simulated field profile in Fig. 4d. The robustness of the topological corner states is examined numerically by simulating the frequency-domain response of the corner states and studying the frequency stability against deformations and defects (see Supplementary Information). Our numerical simulations verify that the frequency of the topological corner states are robust against certain types of disorders and deformations.

## CONCLUSION AND DISCUSSIONS

Exploiting surface-wave PCs, we realized experimentally a higher-order photonic topological state with topological edge states and corner states. The concurrent emergence of topological edge states within the bulk band gap and corner states within the edge band gap demonstrates the higher-order topology in our PCs. In particular, the strongly confined topological corner states provide an efficient approach toward scalable integration of degenerate cavity modes in photonic chips. Besides, our surface-wave PCs offer a versatile platform for the realization of higher-order topology through multiple Bragg scatterings which can yield photonic bandgaps much larger than those in the coupled optical waveguide arrays or other analogs with perturbative tight-binding couplings. The surface-wave PC architecture can also be extended to higher frequency regimes. Our study thus opens a pathway toward higher-order photonic topology with large bandwidths for future photonic science and applications.

## Acknowledgments

This work was sponsored by the National Natural Science Foundation of China under Grants No. 61625502, No. 61574127, No. 61601408, No. 61775193, No. 11704332, No. 11675116, and No. 61801426, the ZJNSF under Grant No. LY17F010008, the Top-Notch Young Talents Program of China, the Fundamental Research Funds for the Central Universities, the Innovation Joint Research Center for Cyber-Physical-Society System, and the Jiangsu province distinguished professor funding.

## Competing Financial Interests

The authors declare no competing financial interests.

## Data and materials availability

All data are presented in the article and Supplementary Materials. Please direct all inquiries to the corresponding author.

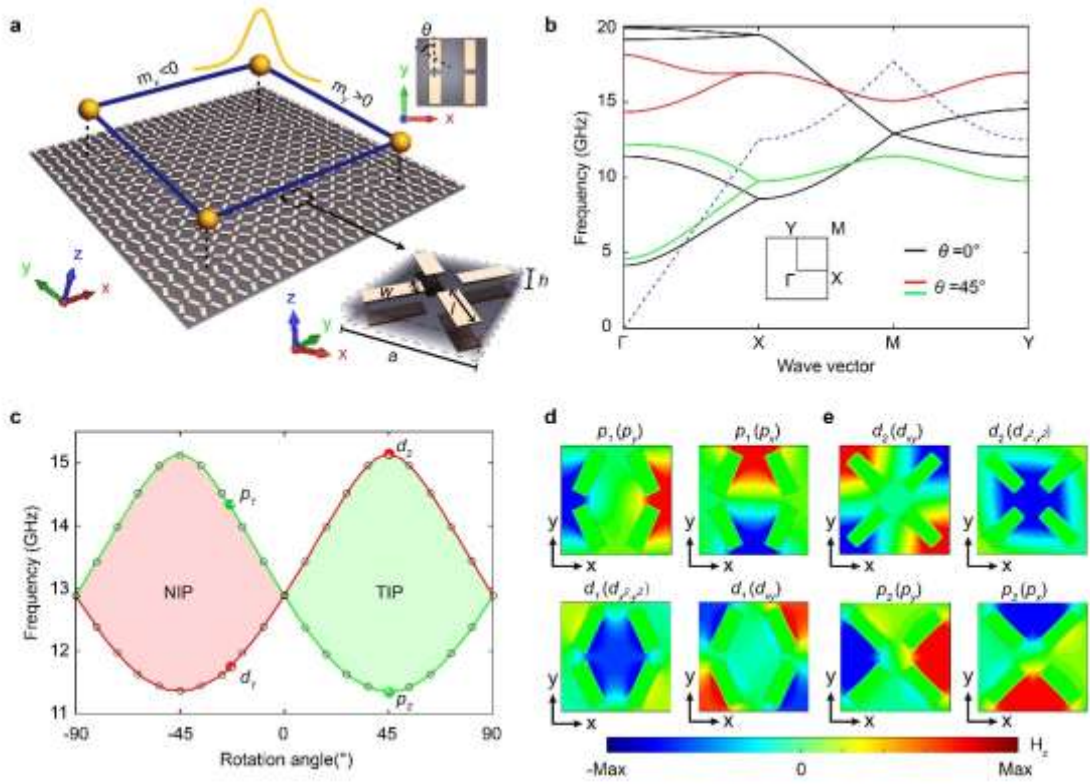

**Figure 1.** Second-order PTI based on surface-wave PCs and its topological transitions. (a) Schematic of topological corner states localized at the corner of the second-order PTI. Based on the Jackiw-Rebbi mechanism, the opposite signs of the Dirac masses along the *x*- and *y*- interfaces lead to the topological corner states. The upper inset shows the top view of the unit cell with rotation angle $\theta$ =0°. The lower inset indicates the unit cell of the designed second-order PTI with $\theta$ =45°. The structure parameters are $h$ =2mm, $w$ =1.92mm, $l$ = 5.04mm, and $a$ =12 mm, respectively. The dielectric substrate has a relative permittivity of 3. (b) Photonic band structures for $\theta$ =0° (black curves) and $\theta$ =45° (red and green curves), respectively. The blue dashed curve represents the light-line in the air. The inset represents the Brillouin zone. (c) Topological phases and the evolution of the photonic band edges at the M point with the rotation angle $\theta$. The green and red curves represent the doubly-degenerate *p* (dipole) and *d* (quadrupole) modes at the M point, respectively. (d) and (e) Magnetic field profiles of the four eigenstates at the M point when $\theta$ =-25°(marked as $p_1$ and $d_1$) and 45°(marked as $d_2$ and $p_2$), respectively.

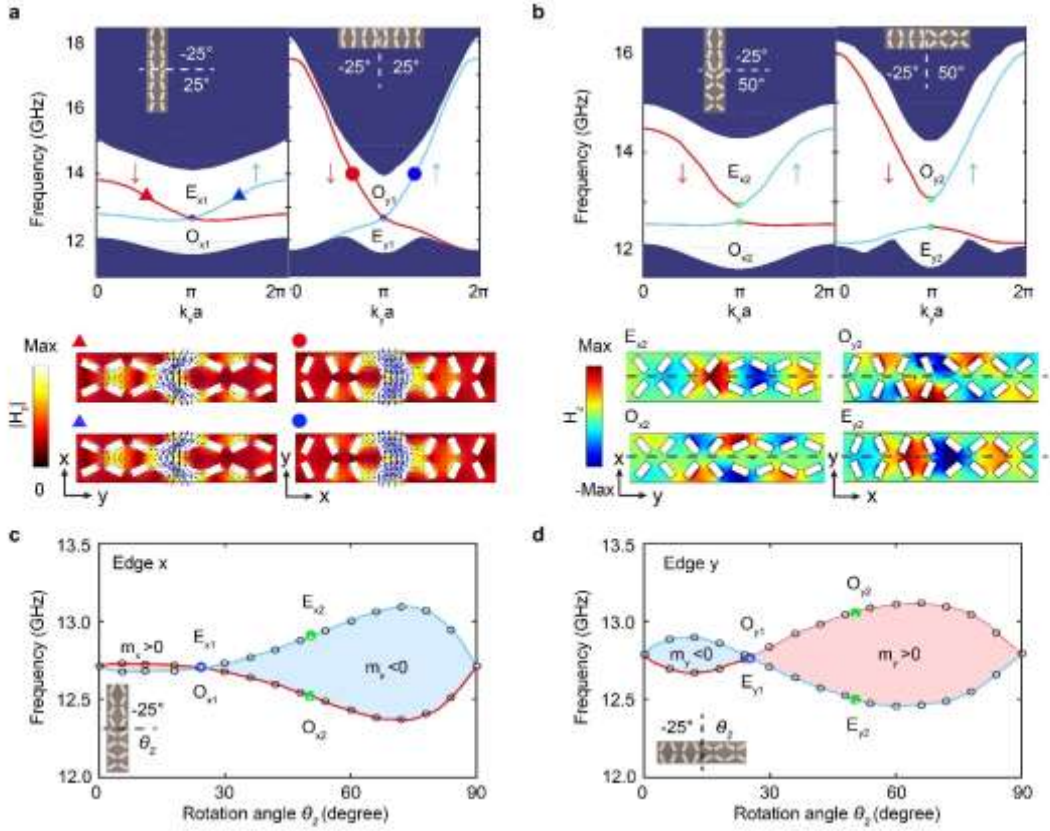

**Figure 2.** Edge states and their topological transitions in the surface-wave PCs. (a) Edge states for the *x*- and *y*-interfaces between the NIP ($\theta_1$ =-25°) and TIP ($\theta_2$ =25°) PCs, respectively. Here, the curves and blue regions denote the dispersions of edge and bulk photonic states, respectively. The blue (red) curves represent the pseudospin-up ↑ (pseudospin-down ↓) edge states. The upper insets indicate the schematics of the supercell. The lower panels represent the field profiles of the edge states labeled by the colored triangles/circles in the dispersions. The Poynting vectors (the blue arrows) indicate the finite orbital angular momenta for the edge states. (b) Edge states at the *x*- and *y*- interfaces between the NIP ($\theta_1$ =-25°) and TIP ($\theta_2$ =50°) PCs, respectively. Here, the lower panels represent the magnetic field distributions of the edge states at $k =\pi/a$. The black dashed lines indicate the symmetric axes. (c) and (d) Topological transitions and Dirac masses of the edge states along *x*- and *y*- interfaces as functions of the rotation angle $\theta_2$, with $\theta_1$ fixed to -25°. The blue (red) curves correspond to the even- (odd-) edge modes at $k=\pi/a$. The blue (green) dots represent the cases of $\theta_2$ =25° (50°). The insets show the schematics of the interface structures.

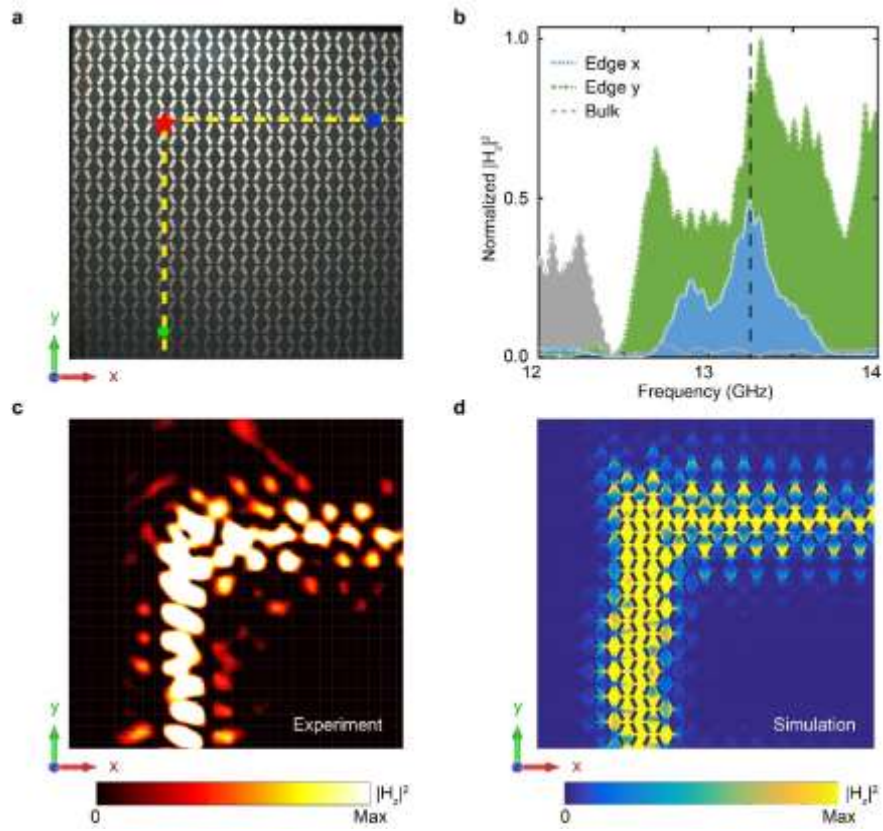

**Figure 3.** Experimental observation of gapless topological edge states in the surface-wave PCs. (a) Perspective-view photograph of the sample (only the upper-left quarter of the structure is shown), composed of the PC with $\theta_2 = 25°$ (TIP, in the lower-right side of the yellow dashed lines) and the PC with $\theta_1 = -25°$ (NIP, in the other region). The red star represents the location of the point source. The green and blue dots denote the locations of the probes. (b) Normalized magnetic field intensity $|H_z|^2$ at the two edge probes (the blue and green regions for the *x*- and *y*-edges, respectively) and at the bulk probe (located at the center of the sample). (c) and (d) Measured and simulated magnetic field intensity distribution $|H_z|^2$ over the sample at 13.24 GHz (marked by the black dashed line in Fig. 3b) excitation, respectively.

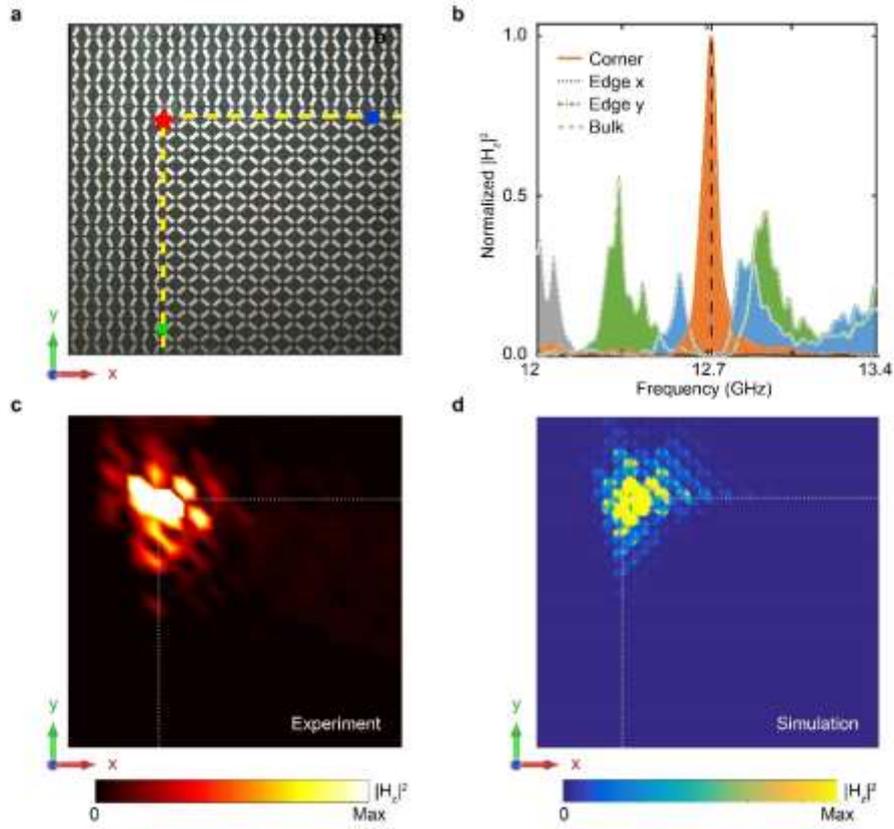

**Figure 4. Experimental observation of topological corner states in surface-wave PCs.** (a) Perspective-view photograph of the experimental sample (only the upper-left quarter of the structure is shown), composed of a PC with $\theta_2 = 50°$ (TIP) (at the lower-right side of the yellow dashed lines), surrounded by the PC with $\theta_1 = -25°$ (NIP). The red star represents the location of the point source. The green and blue dots denote the locations of the probes. (b) Normalized magnetic field intensity $|H_z|^2$ at the two edge probes (the blue and green regions for the $x$- and $y$-edges, respectively) and at the bulk probe (located at the center of the sample). (c) and (d) Measured and simulated magnetic field intensity distribution $|H_z|^2$ over the sample at 12.71 GHz (marked by the black dashed line in Fig. 3b) excitation, respectively. The interface between the TIP and the NIP PCs is labeled by the white dashed lines.